\documentclass[preprint,12pt,longnamesfirst,epsf]{aastex}

\begin{document}

\title{Hot Gas in the Circumstellar Bubble S\,308}

\author{You-Hua Chu\altaffilmark{1}, Mart\'{\i}n A.\ Guerrero\altaffilmark{1,2},
 Robert A. Gruendl\altaffilmark{1}, Guillermo 
Garc\'{\i}a-Segura\altaffilmark{3}, Heinrich J.\ Wendker\altaffilmark{4}}
\altaffiltext{1}{Astronomy Department, University of Illinois, 
        1002 W. Green Street, Urbana, IL 61801;
        chu@astro.uiuc.edu, mar@astro.uiuc.edu, 
        gruendl@astro.uiuc.edu}
\altaffiltext{2}{Now at Instituto de Astrof\'{\i}sica de Andaluc\'{\i}a 
  (CSIC), Spain; mar@iaa.es}
\altaffiltext{3}{Instituto de Astronom\'{\i}a-UNAM, Apartado Postal 877, 
  Ensenada, 22800 Baja California, M\'exico; ggs@astrosen.unam.mx}
\altaffiltext{4}{Hamburger Sternwarte, Gojenbergsweg 112, Hamburg, 
  D-21029 Germany; hjwendker@hs.uni-hamburg.de}


\begin{abstract}

S\,308 is a circumstellar bubble blown by the WN4 star HD\,50896.
It is one of the only two single-star bubbles that show detectable 
diffuse X-ray emission.  
We have obtained {\it XMM-Newton} EPIC observations of the northwest
quadrant of S\,308.
The diffuse X-ray emission shows a limb-brightened morphology,
with a clear gap extending from the outer edge of the diffuse X-ray 
emission to the outer rim of the nebular shell.
The X-ray spectrum of the diffuse emission is very soft, and is well
fitted by an optically thin plasma model for a N-enriched plasma at 
temperatures of $\sim 1.1\times10^6$ K.
A hotter gas component may exist but its temperature is not well
constrained as it contributes less than 6\% of the observed X-ray flux.
The total X-ray luminosity of S\,308, extrapolated from the
bright northwest quadrant, is $\le (1.2\pm0.5)\times10^{34}$ ergs 
s$^{-1}$.
We have used the observed bubble dynamics and the physical parameters
of the hot interior gas of S\,308 in conjunction with the circumstellar
bubble model of \citet{GM95} to demonstrate that the X-ray-emitting gas
must be dominated by mixed-in nebular material.

\end{abstract}  

\keywords{ISM: bubbles --- ISM: individual (S\,308) --- stars: winds 
--- stars: Wolf-Rayet --- stars: individual (HD\,50896) --- 
X-rays: individual (S\,308)}

\section{Introduction}

Fast winds from massive stars can interact with their ambient medium 
and blow bubbles. 
The physical structure of a bubble in a homogeneous interstellar medium 
(ISM) has been modeled by \citet{Wetal77}.
In their model of a pressure-driven bubble, the shocked fast wind 
reaches temperatures of 10$^7$--10$^8$ K and forms a contact 
discontinuity with the cool (10$^4$ K), dense shell of swept-up
ambient medium.  
Heat conduction and mass evaporation across this interface lowers the 
temperature and raises the density of the hot interior gas, producing 
10$^5$--10$^6$ K gas that can be observed at ultraviolet and X-ray
wavelengths.

Of all known bubbles blown by single massive stars, only two show
detectable diffuse X-ray emission: NGC\,6888 and S\,308 
\citep{Boc88,WWW94,Wri99}.
$ROSAT$ PSPC and $ASCA$ SIS observations of NGC\,6888 showed a 
limb-brightened X-ray morphology and spectra that were best fitted by
a dominant component at 1.5$\times$10$^6$ K and a weaker component at
8$\times$10$^6$ K \citep{WWW94,Wetal98,Wetal03}.
$ROSAT$ PSPC observations of S\,308 detected X-ray emission near the 
shell rim, but detailed analysis was hampered by the low surface 
brightness and the obstruction of the PSPC window support structure 
projected near the shell rim \citep{Wri99}.

Both NGC\,6888 and S\,308 have Wolf-Rayet (WR) central stars and show 
enhanced N and He abundances in the nebulae, suggesting that they
consist of stellar material ejected by the central stars during the 
red supergiant (RSG) phase \citep{Eetal92}.
If the RSG mass loss rate was constant, the density of the circumstellar
material would fall off with a $r^{-2}$ dependence, where $r$ is the 
radius.
The formation and evolution of a bubble blown in such a circumstellar 
medium has been calculated analytically by \citet{GM95}, and numerically
simulated by \citet{GLM96} specifically for WR central stars with RSG
progenitors.

\citet{Wetal98,Wetal03} modeled the diffuse X-ray emission from NGC\,6888 
by incorporating Weaver et al.'s (1977) heat conduction in the analytical 
solution of \citet{GM95}, and found that the model could not explain 
simultaneously the observed limb-brightened morphology, surface
brightness, and the presence of a high temperature component.
By reducing the efficiency of heat conduction at the interface, it 
was possible to reproduce the limb-brightened morphology but not the 
surface brightness. 

The recently launched {\sl XMM-Newton Observatory} has unprecedented 
sensitivity in the soft X-ray band, and thus provides an excellent 
opportunity to observe the hot gas in bubble interiors.  
We have obtained {\sl XMM-Newton} observations of S\,308.
This paper reports our analysis of this new X-ray observation and 
discusses the implications of our results on the physical structure
of S\,308's hot interior.

\section{XMM-Newton Observations of S\,308}

S\,308 was observed with the {\sl XMM-Newton Observatory} on 2001 
October 23-24, in Revolution 343, using the EPIC/MOS1, EPIC/MOS2, 
and EPIC/pn CCD cameras (Observation ID 79570201).
The two EPIC/MOS cameras were operated in the full-frame mode for a 
total exposure time of 47.0 ks, and the EPIC/pn camera in the extended
full-frame mode for 43.0 ks.
For all observations, the medium optical blocking filter was used.  
The angular resolution of the EPIC cameras at energies below 1 keV is 
$\sim$6\arcsec, and the energy resolutions of the EPIC/MOS and EPIC/pn
CCDs are 60 eV and 100 eV at 1 keV, respectively \citep{D99,Stetal01}.  

We received the {\sl XMM} pipeline products, and processed them
further using the {\sl XMM-Newton} Science Analysis Software (SAS 
version 5.3.3) and the calibration files from the Calibration Access 
Layer available on 2002 October 22.  
The event files were screened to eliminate bad events, such as those 
due to charged particles.
For the EPIC/MOS observations, only events with CCD patterns 0--12 
(similar to {\sl ASCA} grades 0--4) were selected;  
for the EPIC/pn observation, only events with CCD pattern 0 (single 
pixel events) were selected.  
As the background directly affects the detectability of diffuse X-ray 
emission, periods of high background have to be excluded.
We assessed the background by binning the data over 100~s time
intervals for each instrument in the 10--12 keV energy band; the
background was considered high if the count rates were $\geq$0.3 
cnts~s$^{-1}$ for the EPIC/MOS or $\geq$1.5 cnts~s$^{-1}$ for the 
EPIC/pn.
Unfortunately, 15 ks after the observation of S\,308 started, a 
high-background flare occurred and persisted throughout the rest of the
observation.
The useful exposure times were thus reduced to 22.7 ks and 16.9 ks for 
the EPIC/MOS and EPIC/pn observations, respectively.

\section{Spatial Distribution of Diffuse X-ray Emission from S\,308}

We have observed only the bright northwest quadrant of S\,308 
because the field-of-view of the EPIC cameras is smaller than 
the angular extent of the bubble.
To produce an X-ray image of S\,308, we first extract images in the 
0.25--1.15 keV energy band from the EPIC/pn and EPIC/MOS observations
individually, then mosaic these three images together to increase the 
S/N ratio and to reduce the null exposure in the gaps between CCDs.
This energy band includes all photons detected from S\,308, as shown
later in \S4.
The images are sampled with a pixel size of 2\arcsec, which is adequate 
for the resolution of EPIC cameras.
The mosaicked raw EPIC image is presented in Fig.~1a.  
This image is adaptively smoothed with Gaussian profiles of FWHM ranging
from 2\arcsec\ to 50\arcsec, and divided by a normalized exposure map to
remove the telescope and instrument sensitivity variations across the field.
The resultant image is presented in grey scale with contours in Fig.~1b.

Diffuse X-ray emission from S\,308 is clearly detected, and a 
limb-brightened morphology is seen.
To compare the distribution of diffuse X-ray emission to the cool nebular
shell, we present an [\ion{O}{3}] $\lambda$5007 line image of S\,308 from
\citet{Getal00} in Fig.~1c, and plot the X-ray contours extracted from the 
smoothed EPIC image over the [\ion{O}{3}] image in Fig.~1d.
The diffuse X-ray emission is completely interior to the optical shell.
An apparent gap exists between the outer edge of the diffuse X-ray 
emission and the outer rim of the optical shell.
Close inspection shows bright optical filaments delineating the 
brightest X-ray-emitting regions.
It is not clear whether the gap between the outer edge of the X-ray 
and optical emission regions is caused by a lack of X-ray emission or
additional absorption from the nebular shell.

The optical shell of S\,308 shows a protrusion in the northwest corner,
suggesting a blowout.
This is a very mild case of blowout because its  X-ray edge recedes the 
farthest from the optical shell edge, unlike the superbubble N44's blowout 
where the diffuse X-ray emission extends beyond the optical shell 
\citep{Cetal93}. 

\section{Physical Properties of the Hot Gas in S\,308}

The physical properties of the hot gas in S\,308 can be
determined by modeling the observed X-ray spectrum.
We have extracted spectra from the event files of the
three EPIC cameras separately, using a source aperture
outlined in Fig.~1a, and a background aperture exterior 
to S\,308 but with a similar area, $\sim$210 arcmin$^2$.  
Small regions encompassing X-ray point sources are excised from 
the source and background apertures.
For illustration, regions around the point sources within the 
source aperture are marked in Fig.~1a.
The background-subtracted, vignetting-corrected spectra binned 
by a constant channel width of 15 eV are displayed in Fig.~2.
The EPIC/pn and EPIC/MOS spectra show similar shapes, but the 
EPIC/pn camera is more sensitive and has detected a larger number
of counts, 7,900 $\pm$ 150 cnts.
We have thus concentrated on the EPIC/pn spectrum for further 
analysis and discussion.

The EPIC/pn spectrum of S\,308 is very soft, peaking at the He-like
triplet of \ion{N}{6} at 0.42--0.43 keV and declining sharply toward 
higher energies.
Above 0.7 keV, the spectrum shows only possibly weak emission at 
$\sim$0.92 keV, corresponding to the He-like triplet of \ion{Ne}{9}; 
no emission is detected beyond 1.0 keV.
The presence of line features suggests thermal plasma emission;
therefore, we have used the MEKAL optically-thin plasma emission 
model \citep{KM93,LOG95} to simulate the X-ray spectrum of S\,308.
The observed spectrum is simulated by passing the MEKAL model
spectrum through an interstellar absorption with cross-sections
from \citet{BM92} and convolving it with the EPIC/pn response matrices.
The simulated spectra are compared to the observed spectrum in
the 0.3--2.0 keV range.
The best-fit model, judged by the $\chi^2$ statistics, gives
the plasma temperature, volume emission measure, and 
foreground absorption.  

The input parameters for the spectral fitting include the 
temperature and abundances of the emitting plasma, and the column 
density and abundances of the foreground absorbing medium.
It is impractical to allow all these parameters to vary freely in
the spectral fits because the spectral resolution is limited and
the number of photons detected is modest.
We have therefore used independently determined information to 
constrain the model parameters and fix them whenever possible.
First, we adopt the solar abundances for the foreground
absorbing material, since S\,308 is at a distance of $1.5\pm0.3$
kpc.\footnote{Several distances to HD\,50896 have been reported,
ranging from 0.6\rlap{$^{+0.4}$}{$_{-0.2}$} kpc measured by 
{\it Hipparcos} \citep{Petal97} to $\ge$1.8 kpc determined from 
interstellar absorption lines \citep{HS95}.  We have critically
examined the data from the latter work and conclude that the 
interstellar absorption convincingly demonstrates that HD\,50896 
cannot be closer than 1 kpc, ruling out the near distance by
{\it Hipparcos}.  The systemic velocity of S\,308,
derived from the average velocity near the nebular center, and
the average velocities of the neighboring \ion{H}{2} regions S\,303 
and S\,304 are all near $V_{\rm LSR}$ = $20\pm2$ km~s$^{-1}$
\citep{Cetal82}.  The kinematic distance estimated from this
radial velocity and the Galactic rotation is $1.5\pm0.2$ kpc.
The photometric distance to HD\,50896 is $1.0\pm0.2$ kpc
\citep{vdH01}.  Combining the latter three distances, we 
decide to adopt a distance of $1.5\pm0.3$ kpc to HD\,50896.}
Next we consider the abundances of the X-ray-emitting plasma.  
While the hot gas contains shocked fast stellar wind, its mass
may be dominated by nebular material evaporated across the
conduction front or ablated by the fast stellar wind 
\citep{Wetal77,Petal01a,Petal01b}.
Thus, we start the spectral fits using the nebular abundances
of S\,308.
For He, O, N, and Ne, we adopt the abundances determined from 
optical spectrophotometry by \citet{Eetal92}.
For Mg and Fe, we assume that their abundances relative to 
the solar values are tied to that of O.
The C abundance of the nebula is unknown, although the N/C 
abundance ratio of the fast wind has been determined to be
14 within a factor of 3 uncertainty \citep{H88}.
We adopt a 0.1 solar abundance for C, which is somewhat 
arbitrary but compatible with the stellar wind abundances.
The final abundances of C, N, O, Ne, Mg, and Fe we
have adopted for the MEKAL model are 0.1, 1.6, 0.13, 0.22, 
0.13, and 0.13 times the solar values \citep{AG89}, 
respectively.

Using the above nebular abundances for single-temperature 
MEKAL models, it is possible to reproduce the bulk spectral
shape with a plasma temperature of $kT \sim 0.1$ keV (or
$T \sim 1.1\times10^6$ K).
To improve the spectral fits and to search for the possible 
existence of hotter gas expected from a shocked fast wind, 
we have also fit the observed spectrum with two temperature 
components.
Only a small improvement is achieved, as the reduced $\chi^2$ 
of the single-temperature and two-temperature best fits are
$\chi^2/{\rm DoF}$ = 1.12 and 1.02, respectively.
The best-fit single-temperature model and the low-temperature 
component of the best-fit two-temperature model are virtually
indistinguishable from each other; thus only the
best-fit two-temperature MEKAL model is overplotted on the
EPIC/pn spectrum in Fig.~3.
The absorption column density is
$N_{\rm H} =$ 1.1$\times$10$^{21}$~cm$^{-2}$;
the temperatures, normalization factors, 
observed fluxes, and unabsorbed fluxes of these two temperature
components are summarized in Table 1.
The low temperature component has an accurate temperature and
dominates the X-ray emission, while the high temperature component
is not well constrained in temperature because it is weak and
contributes only 6\% of the observed flux, or 1.5\% of the 
unabsorbed flux.

We have also tried spectral fits using MEKAL models with two 
temperature components and freely varying abundances, and find
that the EPIC/pn spectrum can be reasonably well described by
temperature components at $kT_1 = 0.1$ keV and $kT_2 = 0.7$ keV 
with N, O, and C abundances at 5.6, 0.37, and 0.3 times solar
values, respectively.
These abundances are 3--4 times as high, but the N/C ratio 
($\sim$18 times the solar value) and the temperatures of the two 
thermal components are similar to those of our best-fit model 
with fixed nebular abundances.
Furthermore, both fixed abundance and freely varying abundance 
model fits show that the low-temperature component dominates 
the observed X-ray flux.
For a $1\times10^6$ K plasma with collisional ionization 
equilibrium, N and C exist mainly in He-like ions, and thus the 
N/C abundance ratio can be constrained by the relative strengths 
of \ion{N}{6} lines at 0.43 keV and \ion{C}{5} lines at 0.3 keV, 
or the spectral shape below 0.5 keV. 
The absolute abundances, on the other hand, are not well constrained
because the spectral resolution does not allow unambiguous separation
of line emission and bremsstrahlung emission.
Therefore, we conclude that the diffuse X-ray emission of S\,308 
originates mostly from N-enriched gas at a temperature of 
$\sim 1.1\times10^6$ K.
This temperature is noticeably lower than the plasma temperatures 
observed in planetary nebulae or superbubbles \citep{CGGWK01,GGC02,
Ketal00,Ketal01,Detal03,Tetal03}.

Finally, we have compared our EPIC/pn spectrum of S\,308 to 
Wrigge's (1999) best-fit model for the {\it ROSAT} PSPC spectrum.
Wrigge's model consists of two components at temperatures of
$(1.5\pm0.1)\times10^6$ K and $(2.8\pm0.4)\times10^7$ K, with the 
high-temperature component contributing about 1/3 of the total 
X-ray flux.
Our EPIC/pn spectrum clearly rules out the presence of this 
high-temperature component, which may be caused by a combination 
of a low S/N in the PSPC data and contaminations from the numerous 
point sources projected within the boundary of S\,308. 
These point sources are resolved by the {\it XMM-Newton} EPIC 
cameras but not the {\it ROSAT} PSPC detector.

\section{Discussion}

\subsection{Dynamical Parameters of the S\,308 Bubble}

S\,308 is a bubble blown by the WN4 star HD\,50896 \citep{JH65}
in a circumstellar medium that was ejected by its progenitor
during the RSG phase \citep{Eetal92}.
At a distance of $1.5\pm0.3$ kpc, its 20$'$ radius corresponds to 
$9\pm2$ pc.
The nebular shell of S\,308 is photoionized, but additional shock
heating is present as the nebular [\ion{O}{3}]$\lambda$5007/H$\beta$ 
ratios are as high as $\sim$20 \citep{Eetal92}.
The presence of shocks is also evidenced in the nebular
morphology at the shell rim where the [\ion{O}{3}] emission 
leads the H$\alpha$ emission by 16--20$''$ \citep{Getal00}.
The S\,308 shell resides inside an \ion{H}{1} cavity evacuated by 
the stellar wind of HD\,50896's main-sequence progenitor, with 
the optical nebulae S\,303 and S\,304 \citep[Fig.~1a of][]{Cetal82}
as part of the fossil swept-up interstellar shell \citep{AC96}.
The overall nebular structure and environment of S\,308 agree 
well with the expectations of Garc\'{\i}a-Segura et al.'s 
(1996) models for WR bubbles.

Assuming that the RSG progenitor had a constant mass loss rate
$\dot M_{\rm RSG}$ and a constant wind velocity $V_{\rm RSG}$,
the circumstellar medium of HD\,50896 would have a radial 
density profile $\propto r^{-2}$.  
Assuming further that the WR star has a constant mass loss
rate $\dot M$ and a constant terminal wind velocity $V_\infty$,
the bubble expansion velocity will be constant with 
\begin{equation}
V_{\rm exp} = (\dot M V_\infty V_{\rm RSG} / \dot M_{\rm RSG})^{1/2}
\end{equation}
\citep{GM95}.
For a shell expansion velocity of $63\pm3$ km~s$^{-1}$ \citep{Cetal82}
and a radius of $9\pm2$ pc, the dynamic age of S\,308 is thus 
$(1.4\pm0.3)\times10^5$ yr.
The terminal velocity of HD\,50896's fast stellar wind has been 
well measured to be $V_\infty$ = 1,720 km~s$^{-1}$ \citep{PBH90}.
The mass loss rate of HD\,50896's fast wind has been determined
by \citet{NCW98} from IR and radio continuum measurements assuming
a smooth wind or a clumped wind for a distance of 1.8 kpc.
With the distance adjusted from 1.8 kpc to $1.5\pm0.3$ kpc, the 
smooth-wind and clumping-corrected mass loss rates of HD\,50896 are 
$(3.6\pm1.1)\times10^{-5}$ M$_\odot$ yr$^{-1}$ and 
$(1.4\pm0.5)\times10^{-5}$ M$_\odot$ yr$^{-1}$, respectively.
If we assume that the high electron temperature measured in the
nebular shell, $(1.4-1.8)\times10^4$ K \citep{Eetal92}, is 
caused by shock heating, then the shock velocity has to be 
$\ge$30 km~s$^{-1}$.
The RSG wind velocity can be approximated by the difference 
between the shell expansion velocity and the shock velocity,
thus $V_{\rm RSG} \le$ 30 km~s$^{-1}$; from Eq.\ (1) we then
find the RSG mass loss rate to be $\dot M_{\rm RSG} \le 
(4.7\pm1.4)\times10^{-4}$ M$_\odot$~yr$^{-1}$ for the case of
smooth fast wind, or $ \le (1.9\pm0.6)\times10^{-4}$ 
M$_\odot$~yr$^{-1}$ for the case of clumped fast wind.

The nebular shell of S\,308 has a sharp rim, indicating that
it is still surrounded by RSG wind material; however, the RSG wind 
material must not extend much further in radius, as a breakout has 
already occurred at the northwestern part of the shell.
The unperturbed RSG wind material just exterior to the nebular
shell was ejected at least $(2.9\pm0.6)\times10^5$ years ago.
As the fast WR wind started $(1.4\pm0.3)\times10^5$ yr ago, the
RSG wind lasted $(1.5\pm0.3)\times10^5$ yr, and the total RSG 
mass loss would be $\le 70^{+45}_{-30}$~M$_\odot$ and 
$\le 29^{+15}_{-13}$~M$_\odot$ for the cases of smooth and clumped
fast winds, respectively.
Evidently, the former value is unrealistically large for a normal
massive star; thus the total RSG mass loss confirms Nugis et al.'s 
(1998) suggestion that the clumping-corrected mass loss rates of 
WR stars are more accurate.
Therefore, we will use only the clumping-corrected mass loss rate 
for the rest of the discussion.
The physical parameters of the RSG progenitor and the nebular shell 
derived above are summarized in Table 2.
Note that the velocity, mass loss rate, and total amount of mass 
loss of the RSG wind derived from bubble dynamics are all well 
within the canonical ranges of these parameters, and therefore 
provide a very satisfactory consistency check for the bubble dynamics.

\subsection{The Hot Interior of S\,308}

The {\it XMM-Newton} observations of S\,308 show an X-ray spectrum
dominated by emission from a N-enriched medium at a temperature 
of $\sim 1.1\times10^6$ K, with less than 6\% of the observed flux
contributed by hotter gas.
This dominant temperature is almost two orders of magnitude lower 
than the post-shock temperature of the 1,720 km~s$^{-1}$ fast stellar 
wind, $\sim 8.3\times10^7$ K, for a H/He ratio of 0.2 in the WR
wind \citep{NCW98}.
The cooling time scale for such hot gas is much longer than the 
dynamic age of the bubble; therefore, the observed low plasma 
temperature indicates that cool nebular material has been mixed 
into the shocked WR wind to lower the temperature.
This mixing is supported by the limb-brightened X-ray morphology 
of S\,308.

Two known mechanisms can mix nebular material with the hot gas:
mass evaporation due to thermal conduction and mass loading due 
to dynamic ablation.
The former mechanism has been incorporated into bubble models of 
\citet{Wetal77} and \citet{Petal01a}, and the latter mechanism 
has been used by \citet{Petal01b}.
To determine whether evaporation or ablation is responsible 
for the mixing in S\,308, we need to know whether the WR wind is 
shocked and where the shock front is.
In the bubble model of \citet{Wetal77}, the fast stellar wind
encounters an inner, stagnation shock and the shocked stellar 
wind provides the thermal pressure to drive the expansion of 
the nebular shell.
Thermal conduction at the interface between the hot interior gas 
and the cool nebular shell evaporates nebular material into the 
hot interior, but the bulk of the shocked fast wind remains hot 
and occupies most of the volume in the bubble interior.
We may estimate the X-ray emission expected from the ``uncontaminated"
shocked stellar wind of HD\,50896 and compare it to the observed 
diffuse X-ray emission from S\,308.
The WR wind of HD\,50896 has injected $2\pm1$~M$_\odot$ into the bubble
interior during the lifetime of the bubble.
If the shocked fast wind is uniformly distributed within the 
bubble interior (the worst-case scenario for emissivity and 
detectability), the electron density would be 
$0.015\pm0.003$ cm$^{-3}$, for a H/He ratio of 0.2.
The total volume emission measure of this shocked wind at 
$8.3\times10^7$ K can be scaled to determine its expected
normalization factor $A$ (see Table 1 for the definition of
the normalization factor) for the aperture with which our
EPIC spectrum was extracted; we find an expected normalization
factor to be $(1.2\pm0.1)\times10^{-4}$ cm$^{-5}$.
Compared with the two temperature components from the best
spectral fit listed in Table 1, the normalization factor
expected for the X-ray emission from the uncontaminated
shocked wind is fortuitously close to that of the $T_2$
component.
If the hot shocked wind occupies only a fraction of the total 
volume, the expected normalization factor will be higher
and would have been detected by our {\it XMM-Newton} observation 
if its useful exposure time was not shortened by a high
background.
Deeper X-ray observations are needed to place more 
stringent limits on the existence or absence of the shocked 
stellar wind at $8.3\times10^7$ K.

We have used the best-fit spectral model to determine the 
unabsorbed X-ray flux from S\,308, $7.2\times10^{-12}$
ergs cm$^{-2}$ s$^{-1}$ within the 210 arcmin$^2$ aperture
in the 0.25--1.5 keV energy band.
This X-ray flux can be scaled to the entire area covered
by the bubble to estimate the total X-ray luminosity.
As the observed aperture contains the brightest patch of diffuse 
X-ray emission within S\,308, the X-ray luminosity of S\,308 should 
be $\le (1.2\pm0.5)\times10^{34}$ ergs s$^{-1}$.
The rms electron density of the X-ray-emitting gas in S\,308 can 
be determined from the normalization factor $A$ given in Table 1,
$N_{\rm e} = (0.20\pm0.03)\, \epsilon^{-1/2}$ cm$^{-3}$, where 
$\epsilon$ is the volume filling factor of the hot gas.
For $\epsilon$ = 0.5, the rms $N_{\rm e} = 0.28\pm0.04$ cm$^{-3}$
and the total mass of the hot gas is $11\pm5$ M$_\odot$; while
for $\epsilon$ = 0.1, the rms $N_{\rm e} = 0.63\pm0.09$ cm$^{-3}$ and 
the total mass of the hot gas is $5\pm2$ M$_\odot$.
As the limb-brightened morphology of the diffuse X-ray emission 
indicates a thick shell structure, the filling factor must be closer
to 0.5 than 0.1.
Therefore, the hot gas mass is much larger than the total WR wind 
mass that has been injected into the bubble interior, $2\pm1$ M$_\odot$,
suggesting that nebular material contributes to the hot gas mass
and justifying the fixed nebular abundances we used for the spectral
fits in \S4.

Our present {\it XMM-Newton} observations do not detect enough
photons to warrant a spatially resolved spectral analysis of
S\,308. 
Future deeper observations are needed to determine the radial 
temperature structure of the X-ray emitting gas.
A complete mapping of S\,308 is also needed to determine whether
the varying morphology of the optical shell rim coincides with
spatial and physical variations in the adjacent hot, X-ray 
emitting gas in the bubble interior.

\subsection{Interface between the Hot Interior and the Cool Shell}

UV observations in conjunction with X-ray observations
of S\,308 allow us to probe the interface between the hot
gas in the bubble interior and the cool gas in the swept-up
shell.
High-resolution UV spectra of HD\,50896 have detected a
nebular \ion{N}{5} absorption line component associated with 
the approaching side of the S\,308 shell, and its large column
density suggests an overabundance in nitrogen \citep{Betal97}.
As the ionization potential of \ion{N}{4} is 77.5 eV, \ion{N}{5}
must be produced by collisional ionization in gas at temperatures
of a few $\times10^5$ K, which is expected at the interface 
between the hot interior gas and the cool nebular shell of
a bubble.
The detection of \ion{N}{5} absorption from the S\,308 shell
has provided the strongest evidence of the existence of an
interface layer.

Our {\it XMM-Newton} X-ray image of S\,308 reveals the relative 
location of the hot interior gas and the cool nebular shell 
(see Fig.~1d).
The projected outer edge of the diffuse X-ray emission is offset 
from the projected outer edge of the [\ion{O}{3}] emission by 90$''$ 
to over 200$''$, corresponding to 0.5 -- 1.7 pc.
As the [\ion{O}{3}] rim depicts the outer shock associated with
the swept-up shell advancing into the unperturbed RSG wind,
this ``gap'' contains the swept-up shell and the interface layer 
where the temperature drops radially from 10$^6$ K to 10$^4$ K.
The large angular size of this gap makes it the most promising site 
where an interface layer may be spatially resolved and studied 
in detail.
Future long-slit high-dispersion spectra of the H$\alpha$ + 
[\ion{N}{2}] lines are needed to determine kinematically the exact 
width of the swept-up shell.
We have been awarded {\it Far Ultraviolet Spectroscopic Explorer 
(FUSE)} observations of the \ion{O}{6} emission for three positions
in S\,308.
While these {\it FUSE} observations of \ion{O}{6} emission may 
reveal the existence of $3\times10^5$ K gas at the interface, the
relatively large aperture ($30''\times30''$) provides only 
modest spatial resolution.
Future high-resolution, long-slit {\it Hubble Space Telescope} STIS 
UV spectra of the \ion{N}{5} and \ion{C}{4} emission lines at the gap 
are needed to probe the temperature structure in the interface layer.
Only through the analysis of these detailed multi-wavelength 
observations can we finally learn empirically how fast stellar winds 
interact with the ambient medium and test the various theoretical
models of bubbles.

\section{Summary and Conclusions}

We have obtained {\it XMM-Newton} EPIC observations of the northwest
quadrant of the circumstellar bubble S\,308 blown by the WN4 star
HD\,50896.
Diffuse X-ray emission is unambiguously detected, showing a 
limb-brightened morphology with a clear gap extending from the 
outer edge of the diffuse X-ray emission to the outer rim of 
the nebular shell.
The X-ray spectrum of the diffuse emission is very soft, and is well
fitted by an optically thin plasma model for a plasma at temperatures
of $\sim 1.1\times10^6$ K.
For such a low temperature, the N/C abundance ratio is well constrained
by the spectral shape below 0.5 keV; spectral fits with freely varying
abundances and fixed nebular abundances show consistently that the N/C
ratio of the hot gas in S\,308 is 15--20 times the solar value.
A hotter component may exist, but its temperature is not well
constrained, $8^{+17}_{-6} \times10^6$ K, and contributes less
than 6\% of the observed X-ray flux.

We have applied Garc\'{\i}a-Segura \& Mac Low's (1995) analytical model 
for a WR bubble blown in the progenitor's RSG wind to S\,308, using
observed bubble dynamics and stellar wind parameters, and derived the
masses lost by HD\,50896 via the RSG wind and the WR wind.
We have also used the X-ray data to determine the mass of the 
X-ray-emitting gas in the bubble interior.  The results are summarized
in Table 2.
It can be seen that the mass in the hot gas is indeed much larger
than the total mass injected by the WR wind, indicating that the hot
gas is dominated by the nebular material mixed into the bubble interior.
It is not clear whether the mixing is through conductive evaporation
or dynamic ablation.
Deep X-ray observations are needed to detect hotter gas and to resolve
the temperature structure in the bubble interior.
Spatially resolved UV spectroscopic observations of the interface
between the X-ray-emitting gas and the cool nebular gas are also 
needed to probe the physical mechanisms that mix the nebular material
into the hot interior.

\acknowledgments 
This research was supported by the NASA grant NAG 5-10037.

\clearpage

\begin{figure}
\figurenum{1}
\plotone{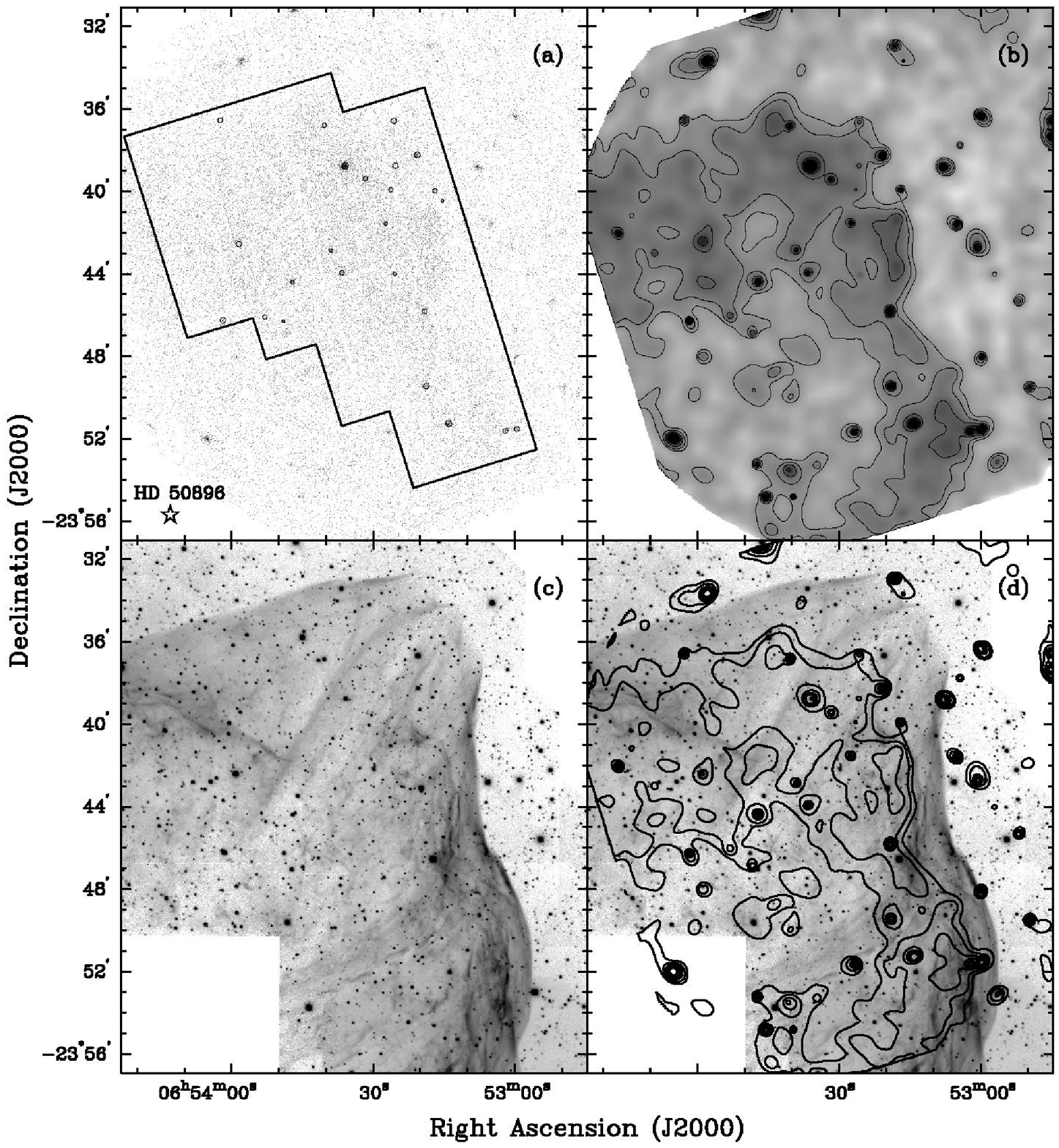}
\caption{(a) {\it XMM-Newton} EPIC image of the northwest
quadrant of S\,308.  The polygon marks the aperture used to
extract spectra of the diffuse X-ray emission.  The small circles
mark the regions centered on the point sources that have been
excluded from the source region when extracting X-ray spectra.
The location of HD\,50896 is marked.
(b) Adaptively smoothed and vignetting corrected X-ray image.
The contours levels are at 3, 5, 10, 20, 30, 40, and 50
$\sigma$ above the background.
(c) [\ion{O}{3}] $\lambda$5007 line image of the northwest 
quadrant of S\,308, taken with the Mount Laguna 1m telescope.
(d) [\ion{O}{3}] image overplotted with X-ray contours.
}
\label{fig1}
\end{figure}

\begin{figure}
\figurenum{2}
\plotone{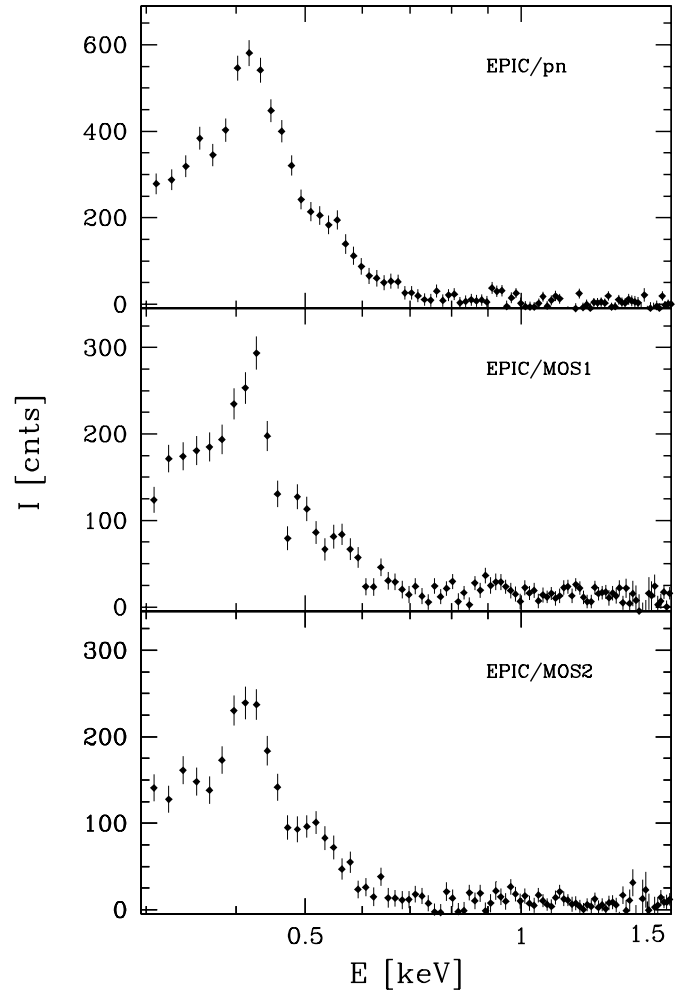}
\caption{{\it XMM-Newton} EPIC spectra extracted from the pn, 
MOS1, and MOS2 cameras, respectively.}
\label{fig2}
\end{figure}

\begin{figure}
\figurenum{3}
\plotone{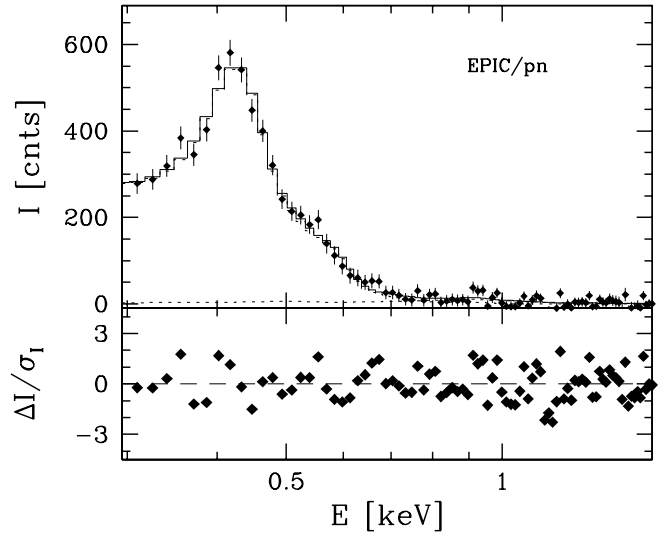}
\caption{{\it XMM-Newton} EPIC/pn spectrum overplotted with the
best-fit model.  The solid curve is the best-fit two-temperature
model, and the dashed curves are the spectra of individual 
components.  The low-temperature component is so dominant that
its spectrum is almost indistinguishable from the sum of both
components; it is also virtually indistinguishable from the 
best-fit of a single-temperature model.
The spectrum of the high-temperature component is the dashed 
curve near the zero line; it contributes 6\% of 
the observed flux, but a negligible number of counts.
The bottom panel shows the residuals.}
\label{fig3}
\end{figure}

\begin{deluxetable}{ccccc}
\tablewidth{0pc}
\tablecaption{Best-fit MEKAL Model for S\,308}
\tablehead{
Components  & $kT$ & $A$\,\tablenotemark{a} &  $f^{\rm obs}_{0.25-1.5}$ & 
$f^{\rm unabs}_{0.25-1.5}$ \\
   &  (keV) &  (cm$^{-5}$)  &  (ergs cm$^{-2}$ s$^{-1}$)  & 
(ergs cm$^{-2}$ s$^{-1}$) 
}
\startdata
$T_1$  &  0.094 $\pm$ 0.009 & $(2.1\pm0.6) \times10^{-2}$ &
$1.0 \times 10^{-12}$   &  $7.2 \times 10^{-12}$  \\
$T_2$  & $0.7^{+1.5}_{-0.5}$ & $(1.3\pm0.8) \times 10^{-4}$ &
$5.8\times10^{-14}$ &  $1.1\times10^{-13}$ \\
\enddata
\tablenotetext{a}{
$A$=$\frac{1.0\times10^{-14}}{4{\pi}d^2}{\int}N_{\rm e}^2 dV$,
where $d$ is the distance, $N_{\rm e}$ is the electron density,
and $V$ is the volume in cgs units.}
\end{deluxetable}

\begin{deluxetable}{llll}
\tablewidth{0pc}
\tablecaption{Physical Parameters of S\,308}
\tablehead{
Parameter     &  Value  & Unit  & Reference
}
\startdata
Distance      &  $1.5\pm0.3$   &  kpc &   \\
Radius        &  $9\pm2$       &   pc  &   \\
Expansion velocity ($V_{\rm exp}$) &  $63\pm3$ &  km s$^{-1}$ &
  \citet{Cetal82}\\
Dynamic age   &  $(1.4\pm0.3)\times10^5$    &   yr  &  \\
Central star  &  HD\,50896    &  & \citet{vdH01} \\
Spectral type &  WN4    &   &  \citet{vdH01} \\
Mass loss rate ($\dot M$) & $(1.4\pm0.5)\times10^{-5}$  &  M$_\odot$ yr$^{-1}$
  & \citet{NCW98}\tablenotemark{a} \\
Terminal wind velocity ($V_\infty$)  &  1,720   &  km s$^{-1}$ 
  & \citet{PBH90} \\
RSG mass loss rate ($\dot M_{\rm RSG}$)  &  $\le (1.9\pm0.6)\times10^{-4}$ & 
 M$_\odot$ yr$^{-1}$  &  \\
RSG wind velocity ($V_{\rm RSG}$)  &  $\le$ 30   & km s$^{-1}$ &  \\
Total RSG wind mass &  $29^{+15}_{-13}$  &  M$_\odot$ \\
Total WR wind mass  &  $2\pm1$   &  M$_\odot$ \\
$N_{\rm e}$ of the X-ray-emitting gas:  &   & \\
($\epsilon=0.5$) &  $0.28\pm0.04$ &  cm$^{-3}$  \\
($\epsilon=0.1$) &  $0.63\pm0.09$ &  cm$^{-3}$  \\
Mass of the X-ray emitting gas:  &   &  \\
($\epsilon=0.5$) &  $11\pm5$  &  M$_\odot$ \\
($\epsilon=0.1$) & $5\pm3$  &  M$_\odot$ \\
\enddata
\tablenotetext{a}{This is their clumping-corrected mass loss rate
with the distance adjusted from their 1.8 kpc to our $1.5\pm0.3$ kpc.}
\end{deluxetable}


\begin{thebibliography}{}

\bibitem[Anders \& Grevesse(1989)]{AG89}
  Anders, E., \& Grevesse, N.\ 1989, Geochim.\ Cosmochim.\
  Acta, 53, 197

\bibitem[Arnal \& Cappa(1996)]{AC96}
Arnal, E.~M., \& Cappa, C.~E.\ 1996, \mnras, 279, 788

\bibitem[Baluci\'nska-Church \& McCammon(1992)]{BM92}
Baluci\'nska-Church, M., \& McCammon, D.\ 1992, \apj, 400, 699

\bibitem[Bochkarev(1988)]{Boc88} Bochkarev, N.\ G.\ 1988,
  Nature,  332, 518

\bibitem[Boroson et al.(1997)]{Betal97} Boroson, B., McCray, R., 
   Clark, C.~O., Slavin, J., Mac Low, M., Chu, Y., \& Van 
   Buren, D.\ 1997, \apj, 478, 638 [Erratum: 1997, \apj, 485, 436]

\bibitem[Chu et al.(2001)]{CGGWK01} Chu, Y.-H., Guerrero, M.~A., 
Gruendl, R.~A., Williams, R.~M., \& Kaler, J.~B.\ 2001, \apjl, 553, L69 

\bibitem[Chu et al.(1993)]{Cetal93} Chu, Y.-H., Mac Low, M.~M., 
   Garc\'{\i}a-Segura, G., Wakker, B., \& Kennicutt, R.~C.\ 1993, 
   \apj, 414, 213 

\bibitem[Chu et al.(1982)]{Cetal82} Chu, Y.-H., Gull, T.~R., Treffers,
  R.~R., Kwitter, K.~B., \& Troland, T.~H.1982, \apj, 254, 562 

\bibitem[Dahlem(1999)]{D99} 
   Dahlem, M.\ 1999, XMM Users' Handbook

\bibitem[Dunne et al.(2003)]{Detal03} Dunne, B.~C., Chu, Y.-H., 
 Chen, C.-H.~R., Lowry, J.~D., Townsley, L., Gruendl, R.~A., Guerrero, 
 M.~A., \& Rosado, M.\ 2003, \apj, 590, 306 

\bibitem[Esteban et al.(1992)]{Eetal92} 
  Esteban, C., V\'{i}lchez, J.~M., Smith, L.~J., \& Clegg, R.~E.~S.\ 
  1992, \aap, 259, 629 

\bibitem[Garc\'{\i}a-Segura et al.(1996)Garc\'{\i}a-Segura, Langer, 
  \& Mac Low]{GLM96} 
  Garc\'{\i}a-Segura, G., Langer, N., \& Mac Low, M.-M.\ 1996, 
  \aap, 316, 133 

\bibitem[Garc\'{\i}a-Segura \& Mac Low(1995)]{GM95} 
  Garc\'{\i}a-Segura, G.,~\& Mac Low, M.\ 1995, \apj, 455, 145 

\bibitem[Gruendl et al.(2000)]{Getal00} 
Gruendl, R.\ A., Chu, Y.-H., Dunne, B.\ C., \& Points, S.\ D.\ 
2000, \aj, 120, 2670 

\bibitem[Guerrero, Gruendl, \& Chu(2002)]{GGC02} Guerrero, 
M.~A., Gruendl, R.~A., \& Chu, Y.-H.\ 2002, \aap, 387, L1 

\bibitem[Hillier(1988)]{H88} Hillier, D.~J.\ 1988, \apj, 327, 822 

\bibitem[Howarth \& Phillips(1986)]{HP86}
Howarth, I.\ D., \& Phillips, A.\ P. 1986, \mnras, 222, 809

\bibitem[Howarth \& Schmutz(1995)]{HS95} Howarth, I.~D.~\& 
Schmutz, W.\ 1995, \aap, 294, 529 

\bibitem[Johnson \& Hogg(1965)]{JH65} Johnson, H.~M., \& 
  Hogg, D.~E.\ 1965, \apj, 142, 1033 

\bibitem[Kaastra \& Mewe(1993)]{KM93}
  Kaastra, J.\ S., \& Mewe, R.\ 1993, Legacy, 3, 16, HEASARC, NASA

\bibitem[Kastner et al.(2000)]{Ketal00} Kastner, 
J.~H., Soker, N., Vrtilek, S.~D., \& Dgani, R.\ 2000, \apjl, 545, L57 

\bibitem[Kastner et al.(2001)Kastner, Vrtilek, \& Soker]{Ketal01} 
Kastner, J.~H., Vrtilek, S.~D., \& Soker, N.\ 2001, \apjl, 550, L189 

\bibitem[Liedahl et al.(1995)Liedahl, Osterheld, \& Goldstein]{LOG95}
  Liedahl, D.\ A., Osterheld, A.\ L., \& Goldstein, W.\ H.\ 1995, 
  \apj, 438, L115

\bibitem[Nugis et al.(1998)Nugis, Crowther, \& Willis]{NCW98}
Nugis, T., Crowther, P.~A., \& Willis, A.~J.\ 1998, \aap, 333, 956

\bibitem[Perryman et al.(1997)]{Petal97} Perryman, M.~A.~C.~et 
al.\ 1997, \aap, 323, L49 

\bibitem[Pittard et al.(2001a)Pittard, Dyson, \& Hartquist]{Petal01a}
 Pittard, J.~M., Dyson, J.~E., \& Hartquist, T.~W.\ 2001a, \aap, 367, 1000

\bibitem[Pittard et al.(2001b)Pittard, Hartquist, \& Dyson]{Petal01b}
 Pittard, J.~M., Hartquist, T.~W., \& Dyson, J.~E.,\ 2001b, \aap, 373, 1043

\bibitem[Prinja et al.(1990)Prinja, Barlow, \& Howarth]{PBH90}
Prinja, R.~K., Barlow, M.~J., \& Howarth, I.~D.\ 1990, \apj, 361, 607

\bibitem[Str\"uder et al.(2001)]{Stetal01}
         Str\"uder, L.\ et al.\ 2001, \aap, 365, 18

\bibitem[Townsley et al.(2003)]{Tetal03}
 Townsley, L.~K., Feigelson, E.~D., Montmerle, T., Broos, P.~S.,
 Chu, Y.-H., Garmire, G.~P.\ 2003, \apj, 593, 874 

\bibitem[van der Hucht(2001)]{vdH01}
van der Hucht, K.~A.\ 2001, New Astronomy Review, 45, 135

\bibitem[Weaver et al.(1977)]{Wetal77} Weaver, R., McCray, R., 
  Castor, J., Shapiro, P., \& Moore, R.\ 1977, \apj, 218, 377 

\bibitem[Wrigge(1999)]{Wri99} Wrigge, M.\ 1999, \aap, 343, 599 

\bibitem[Wrigge et al.(1998)]{Wetal98} Wrigge, M., Chu, Y.-H.,
  Magnier, E.~A., \& Kamata, Y.\ 1998, Lecture Notes 
  in Physics, v.506, Berlin Springer Verlag, 506, 425 

\bibitem[Wrigge et al.(2003)]{Wetal03} Wrigge, M., Chu, Y.-H.,
  Magnier, E.~A., \& Wendker, H.~J.\ 2003, \apj, submitted

\bibitem[Wrigge et al.(1994)Wrigge, Wendker, \& Wisotzki]{WWW94} 
  Wrigge, M., Wendker, H.~J., \& Wisotzki, L.\ 1994, \aap, 286, 219 

\end{thebibliography}
\end{document}